# Erbium-implanted WS$_2$ flakes with room-temperature photon emission at telecom wavelengths

Guadalupe García-Arellano[1], Gabriel I. López Morales[1,4], Zav Shotan[1], Raman Kumar[1], Ben Murdin[3], Cyrus E. Dreyer[4,5], and Carlos A. Meriles[1,2,*]

[1]Department of Physics, CUNY-City College of New York, New York, NY 10031, USA.
[2]CUNY-Graduate Center, New York, NY 10016, USA.
[3]Advanced Technology Institute, University of Surrey, Guildford GU2 7XH, United Kingdom.
[4]Department of Physics and Astronomy, Stony Brook University, Stony Brook, New York, 11794-3800, USA.
[5]Center for Computational Quantum Physics, Flatiron Institute, 162 5th Avenue, New York, New York 10010, USA.

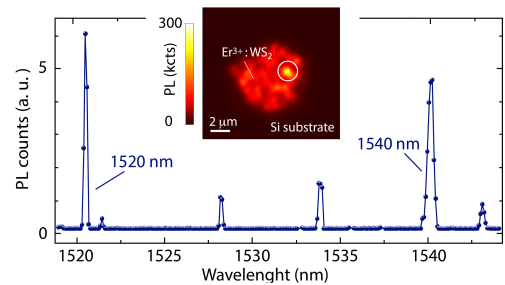

**ABSTRACT:** Optically addressable spin impurities in crystals along with device engineering provide an attractive route to realizing quantum technologies in the solid state, but reconciling disparate emitter and host material constraints for a given target application is often challenging. Rare-earth ions in two-dimensional (2D) materials could mitigate this problem given the atomic-like transitions of the emitters and the versatile nature of van der Waals systems. Here we combine ion implantation, confocal microscopy, and ab-initio calculations to examine the photon emission of Er-doped WS$_2$ flakes. Optical spectroscopy reveals narrow, long-lived photo-luminescence lines in the telecom band, which we activate after low-temperature thermal annealing. Spectroscopic and polarization-selective measurements show a uniform response across the ensemble, while the fluorescence brightness remains mostly unchanged with temperature, suggesting non-radiative relaxation channels are inefficient. Our results create opportunities for novel solid state devices coupling 2D-hosted, telecom-band emitters to photonic heterostructures separately optimized for photon manipulation.

Solid-state quantum registers formed by interacting electron and nuclear spins amenable to high-fidelity state manipulation and readout provide a promising architecture for quantum technologies[1,2]. Quantum emitters in the form of substitutional rare-earth ions have surfaced as an interesting option because electronic screening of the 4$f$ orbitals leads to narrow optical transitions at frequencies relatively insensitive to the crystal host. Since not all materials are equally flexible to device engineering, this feature is advantageous in that it allows the experimenter to separately select the optimal emitter/host system for a given application. With emission in the telecom band[3], Er$^{3+}$ ions are especially attractive for long-distance communication and distributed quantum computing, an interest accentuated by recent demonstrations of sub-diffraction single-shot readout[4], indistinguishable photon generation[5], quantum storage of photon states[6,7], and spectral multiplexing[8].

While most applications to quantum information processing have focused thus far on rare-earth ions in garnets or oxides[9], an intriguing possibility is the use of two-dimensional (2D) hosts[10,11], especially in multilayer form. Unlike monolayers — where emitters are subject to environmental fluctuations difficult to control — these 'quasi-2D' hosts provide bulk-like crystalline environments, which facilitates control of the qubit properties through the use of metamaterial structures[12,13], local strain[14-16], or electric fields[17-19]. Given the sub-wavelength distances involved, 2D-hosted emitters can be near-field coupled to photonic structures explicitly designed for photon manipulation[20].

From a broad palette of 2D systems, tungsten disulfide (WS$_2$) garners particular interest given its chemical stability, high electron mobility[21], and bandgap tunability[22]. All nuclear-spin-active isotopes have low natural abundances suggesting Er ions in this material could serve as long-lived spin qubits[23,24]. Moreover, the non-polar nature of the crystalline lattice should render substitutional color centers robust to noise processes (e.g., moving charges) known to degrade the emitter's optical and spin properties, especially close to the surface[25,26].

Er-doped WS$_2$ nanosheets in the form of composite films have recently been examined as a platform for photon up- and down-conversion[27], infrared (IR) photodetection[27], and photothermal imaging[28]. In the same vein, chemical vapor deposition (CVD) has been shown to yield large-area, Er:WS$_2$ flakes with Er photoluminescence (PL) in the infrared[29,30]. Unfortunately, all reported spectra feature broad emission lines (50–100 nm) pointing to crystalline disorder and/or contributions from non-equivalent Er ions, inadequate for quantum information processing applications.

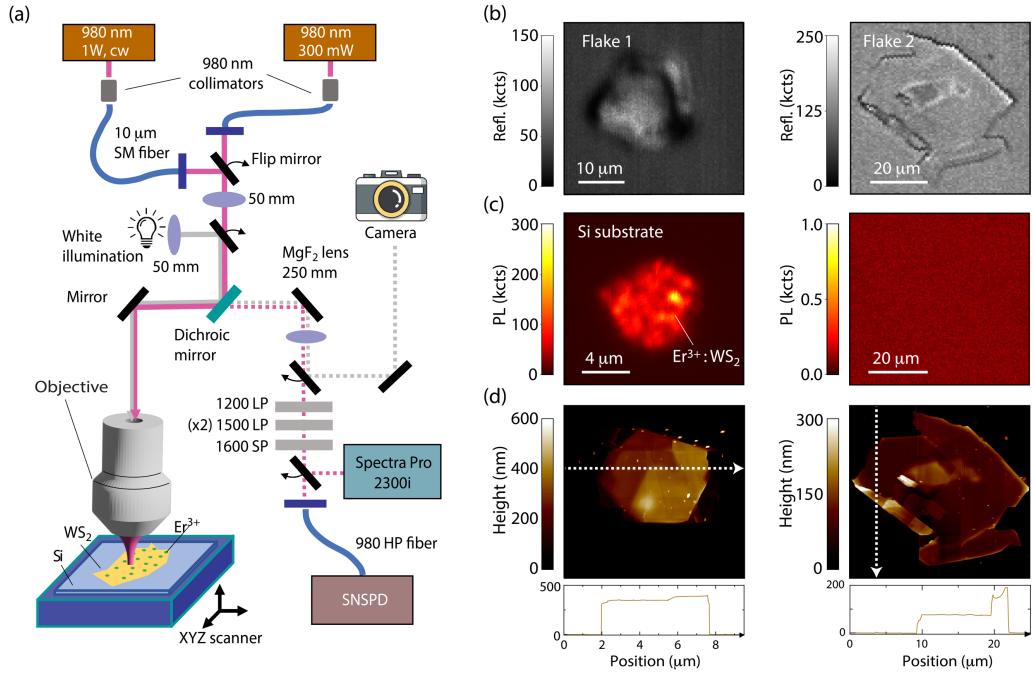

**Figure 1: Infrared fluorescence confocal microscopy of exfoliated WS$_2$.** (a) Schematic of the experimental setup. Depending on the target experiment, we resort to cw or gated 980-nm lasers for excitation; both lasers are linearly polarized. We use three long-pass filters (one at 1200 nm and two at 1500 nm) as well as a short-pass filter (at 1600 nm) to minimize photon contributions away from the telecom bands. (b) Reflection-mode confocal images of select flakes on the silicon substrate. In this modality, we remove all bandpass filters and collect the excitation beam reflection upon proper attenuation. (c) Confocal fluorescence images of the flakes shown in (b). Flake 2 yields no fluorescence, a consequence of its thickness being thinner than the penetration depth of the Er ions during implantation. (d) Atomic force microscopy of the flakes in (b). The lower plots are cross sections of the images along the dashed white lines. All experiments at room temperature. SNSPD: Superconducting nanowire single photon detector. LP: Long-pass filter. SP: Short-pass filter. SM: Single mode fiber. HP: High power fiber.

Here, we articulate confocal microscopy and density functional theory to study the telecom PL stemming from Er emitters in exfoliated WS$_2$ flakes. Variable temperature optical spectroscopy reveals a collection of emission lines featuring sub-nm inhomogeneous linewidths and ms-long lifetimes, as well as a high-degree of linear polarization. Using embedding techniques to properly capture the many-body nature of the ion, we derive emission spectra and polarization plots for substitutional Er$^{3+}$ emitters in qualitative agreement with our observations.

The sample we examine comprises a collection of WS$_2$ flakes exfoliated from a high-purity crystal and transferred onto a Si substrate. In the absence of available CVD reactors[29,30], we resort to broad-area Er ion implantation. Unlike incorporation during growth, ion bombardment introduces lattice damage, which here proved sufficient to quench all Er fluorescence in the as-implanted flakes (possibly due to high vacancy concentration[29]). We circumvented this problem via a soft, 1-hour-long thermal annealing at 400 $^0$C in an Ar environment (see Supplementary Information (SI), Section I). These conditions are seemingly sufficient to improve the crystallinity of WS$_2$ and selectively activate Er emission in the flakes without also triggering fluorescence from ions in the Si substrate[31].

We use a custom-made confocal microscope with excitation at 980 nm (Fig. 1a) to examine flake sets exposed to varying Er implantation doses (see SI, Section I for details). Fig. 1b displays two examples — Flakes 1 and 2, with implantation dose of $10^{14}$ ions/cm$^2$— here imaged optically upon reconfiguring the microscope to monitor the back-reflection of the 980-nm excitation beam. As shown in the fluorescence maps of Fig. 1c, only the first one exhibits telecom-band PL. Our observations indicate the flake thickness — smaller for Flake 2, see Fig. 1d — is the underlying cause. Indeed, a systematic comparison between flakes shows that a thickness of at least 200 nm is necessary to observe Er fluorescence (SI, Section I). For the present implantation energy (75 keV), we find a penetration depth of ~400 nm which, in turn, exposes a large discrepancy with predictions for Er ions extracted from SRIM[32] modeling[33-36].

To investigate the optical response of the WS$_2$-hosted emitters, we now turn our attention to Flake 1 where optical spectroscopy across the 1300–1600 nm range reveals a collection of PL emission peaks featuring sub-nm inhomogeneous linewidths, often limited by the spectrometer resolution (0.1 nm). Qualitatively, we interpret our observations as the result of a cascade process, first involving relaxation from $^4I_{11/2}$ (pumped via

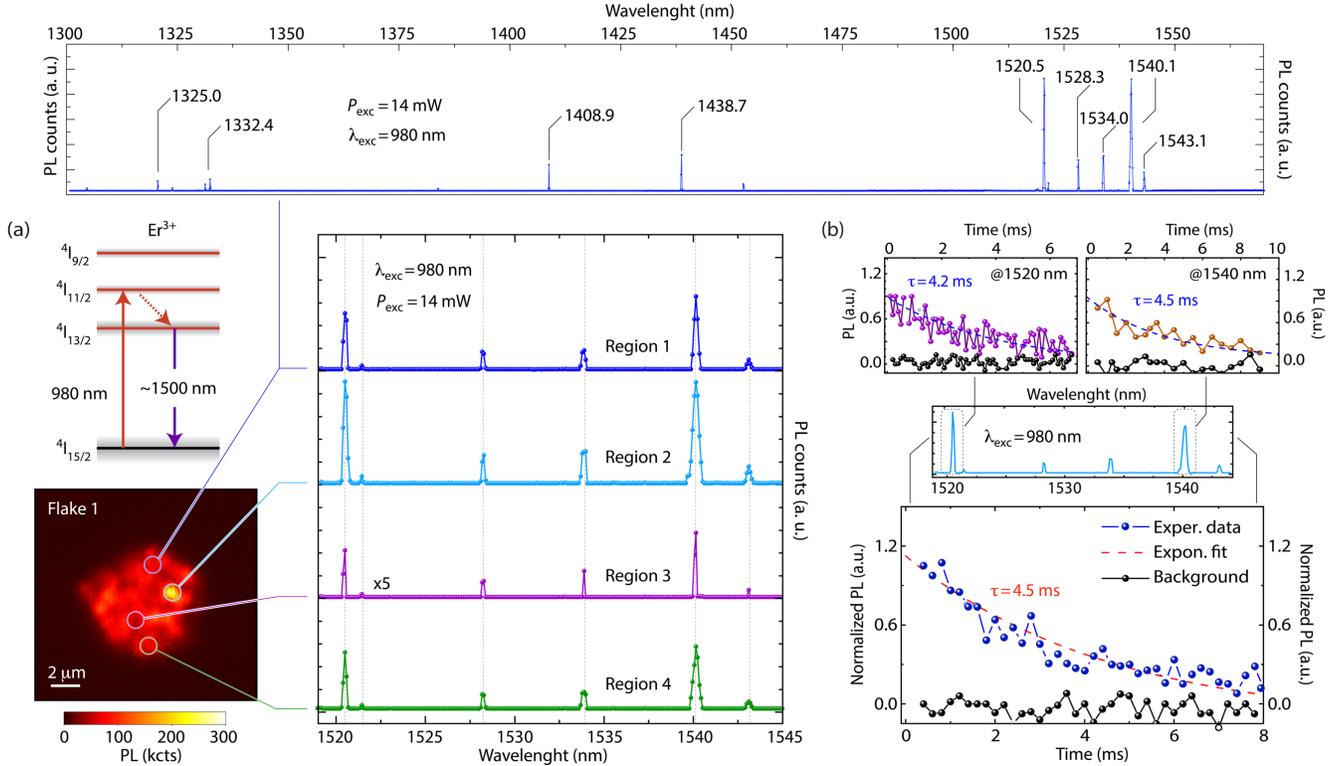

**Figure 2. Optical characterization of Er-implanted WS$_2$ flakes.** (a) Schematic energy diagram (top left) and fluorescence confocal image of Flake 1 (bottom left, reproduced from Fig. 1 for clarity). Colored circles highlight the areas of the flake corresponding to the optical spectra on the top and right-hand side plots. In all cases, we use 14 mW, 980 nm laser excitation; spectra have been displaced vertically for clarity. (b) (Bottom) PL amplitude integrated over a 1500–1600 nm window (blue dots) as a function of the time elapsed after pulsed optical excitation. From an exponential fit (red, dashed line), we derive an excited state lifetime $\tau = 4.5$ ms. Black dots indicate the observed response in the absence of an excitation pulse. (Upper inserts) Same as before but for selective detection at 1520 and 1540 nm (left and right plots, respectively). All experiments at room temperature.

980 nm excitation) into $^4I_{13/2}$, followed by photon emission upon decay to the ground state $^4I_{15/2}$ (Fig. 2a); exchange and crystal field (CF) interactions transform into manifolds each of these high-multiplicity levels, hence leading to several emission lines, as observed.

Spectra from different sites reveal excellent uniformity, suggesting good crystallinity. Note, however, that areas with lower integrated PL intensity tend to exhibit the narrowest lines (see, e.g., spectrum from Region 3 in Fig. 2a), which hints at some residual lattice heterogeneity in sections of the crystal with the highest brightness. Flakes exposed to lower Er implantation doses show comparatively more uniform PL maps (as well as improved conversion efficiency upon annealing), possibly because lower ion concentrations prevent Er clustering[37,38] (SI, Section I).

Optical relaxation in rare-earth ions is typically slow given the shielding of the 4$f$ orbitals by lower-energy, outer-radii 6$s$ and 5$p$ electrons. We confirm this feature in Fig. 2b where we monitor the flake's luminescence as it decays following pulsed optical excitation at 980 nm. We measure a characteristic lifetime $\tau = 4.5 \pm 0.3$ ms, comparable to that seen in other high-quality hosts[9] (SI, Section I). In principle, non-radiative processes could be present although the slow PL decay we measure seems to indicate radiative mechanisms dominate spontaneous relaxation. Observations at temperatures down to 3.5 K — showing little PL change in brightness, frequency, or lifetime, see SI, Section I — confirm this idea.

While optical spectroscopy alone is insufficient to unveil the microscopic structure of the emitters, a substitutional Er$_W$ geometry — where Er occupies the site of a W atom in the WS$_2$ lattice — seems the most plausible given the much smaller atomic radius of S. In fact, preceding density functional theory (DFT) studies indicated Er$_W$ forms a stable point defect with minimum associated lattice distortion[23,24,29]. Formation energies were seen to favor the neutral and negatively charged defect states, though the electronic occupation of Er remained that of its trivalent form across all charge states (excess electrons occupy defect states stemming from dangling bonds). Further, reduced hybridization of the Er$^{3+}$ orbitals with the WS$_2$ lattice resulted in a set of atom-like transitions in the telecom range[23,24], although their accuracy is questionable since DFT alone cannot capture the many-body interactions inherent to rare-earth ions.

To describe the optical properties of this system on a more quantitative footing, we calculate the optical matrix elements between many-body states through a correlated semi-empirical model within the framework of quantum embedding[39-42] (see SI, Section II). We assume the Er defect is in the neutral charge state (Er$_W^0$), though our results are qualitatively similar for the negatively charged defect. We consider two situations: In the first scenario, we assume the lowest CF state of $^4I_{13/2}$ is the only one populated, and transitions occur to all $^4I_{15/2}$ CF states; the second case is one where transitions occur between all $^4I_{13/2}$ and $^4I_{15/2}$ CF states (Fig. 3a and 3b, respectively). The former describes the situation where carriers in the excited $^4I_{13/2}$ manifold fully thermalize before the emission process, whereas the latter is the case where there is some population in all $^4I_{13/2}$ states. Likely, our experiments lie somewhere between these two limit cases, depending on the details of the relaxation process from the $^4I_{11/2}$ states and non-radiative processes within the dense $^4I_{13/2}$ manifold[9,43-47]. In both cases, we see that Er$_W^0$ indeed features bright emission in a region consistent with that observed, but the number of PL lines changes, hence complicating a comparison with experiment.

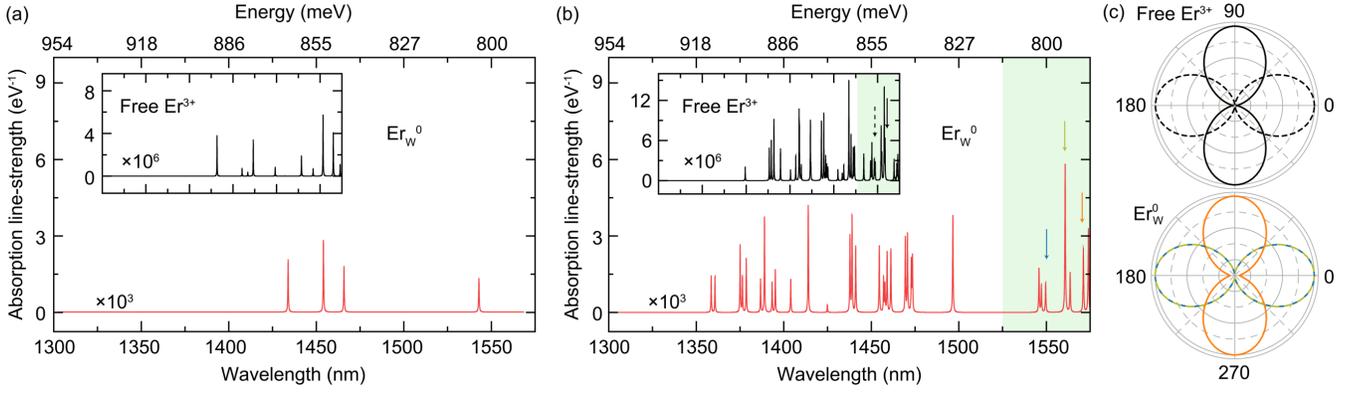

**Figure 3. Calculated $^4I_{15/2} \leftrightarrow {}^4I_{13/2}$ optical transitions of Er$_W$ in WS$_2$.** (a) Absorption line strengths between the lowest-energy crystal-field (CF) state of the $^4I_{15/2}$ manifold (ground state) and all $^4I_{13/2}$ CF states. The inset compares the same for the case of the isolated Er$^{3+}$ ion in free space. (b) Optical transition strengths from all CF states of the $^4I_{15/2}$ manifold to all the CF states of the $^4I_{13/2}$ manifold. Green shaded regions represent the region of focus in the optical experiments of Fig. 2. All lines were convolved with a 0.1-meV-broad Lorentzian for clarity. (c) Polarization dependence of select CF transitions in (b), as highlighted by color-coded arrows. In the top panel of (c), the polarization of two proximal transitions (black arrows in the inset to (b)) are represented by solid and dashed lines to illustrate different degrees of polarization in the transitions of the free ion. The bottom panel of (c) exemplifies the polarization for three CF transitions of the Er$_W^0$ defect that lie close to 1550 nm (color-coded arrows in (b)); note we use a dashed line for the blue trace so as to reveal the underlying green trace.

Our calculations, however, do provide some valuable insights. For example, in Figs. 3a and 3b we plot the results for an isolated Er$^{3+}$ ion in a box with the same dimensions as the defect cell. In this case, the CF splittings derive from unscreened interactions between the charged Er$^{3+}$ ion and its periodic images. The fact that the inter-$f$ transitions in the free Er$^{3+}$ are three orders of magnitude lower than in Er$_W^0$ shows that hybridization with the crystal, and not just a CF interaction, is necessary to activate Er emission. As a matter of fact, the excitation amplitudes of many of the individual $^4I_{15/2} \leftrightarrow {}^4I_{13/2}$ CF transitions in Er$_W^0$ result in radiative lifetimes between 3–10 ms, which fall within those measured (Fig. 2b).

Comparing the calculated optical matrix elements along orthogonal directions in the cell, we find that most of the CF transitions feature a strong degree of polarization. We illustrate this dependence in Fig. 3c where we choose a few transitions near 1540 nm and calculate the angular response of its net transition dipole moment. Note that the strong polarization in the isolated ion suggests this dependence is intrinsic to the $4f \leftrightarrow 4f$ transitions in Er$^{3+}$.

While establishing a one-to-one correspondence between the calculated and measured transitions seems presently unwarranted, experiment can help validate the predicted polarization dependence. As an illustration, Fig. 4 shows data from Flake 3, produced under conditions identical to Flake 1. In agreement with ab-initio modeling, selective imaging of the 1540 nm PL peak reveals a strong polarization dependence, both on excitation and emission (Figs. 4a and 4b). Further, a comparison with polarization plots at 1521 nm (Fig. 4c) shows that the direction of the emission dipole changes by 90 degrees relative to that observed at 1540 nm, qualitatively consistent with the ab-initio predictions in Fig. 3c. Interestingly, we find the excitation and emission dipoles are invariably parallel, even for transitions featuring perpendicular emission polarization (Fig. 4c): Given the off-resonant nature of the excitation (see Fig. 1a), this observation suggests that $^4I_{11/2}$ relaxation into the $^4I_{13/2}$ manifold follows polarization-preserving selection rules. As a whole, our observations and ab-initio calculations correlate well with previous work[47-50] (SI, Sections I and II).

In summary, we combined sample engineering and IR microscopy to demonstrate telecom emission of Er$^{3+}$ in exfoliated WS$_2$. Optical spectroscopy reveals a pattern of narrow PL lines with minimal heterogeneity, an indication of good crystalline quality. The PL brightness and lifetime depend only weakly on temperature, from cryogenic to ambient conditions, comparing favorably to the behavior of Er in other hosts[51]. Quantum embedding calculations of Er$_W$ show that the presence of the WS$_2$ lattice activates nearly forbidden transitions in the Er$^{3+}$ ion core to yield a set of PL lines in the telecom band. Our modeling predicts linearly polarized transition dipoles with transition-dependent, perpendicular orientations, a feature we also observed experimentally.

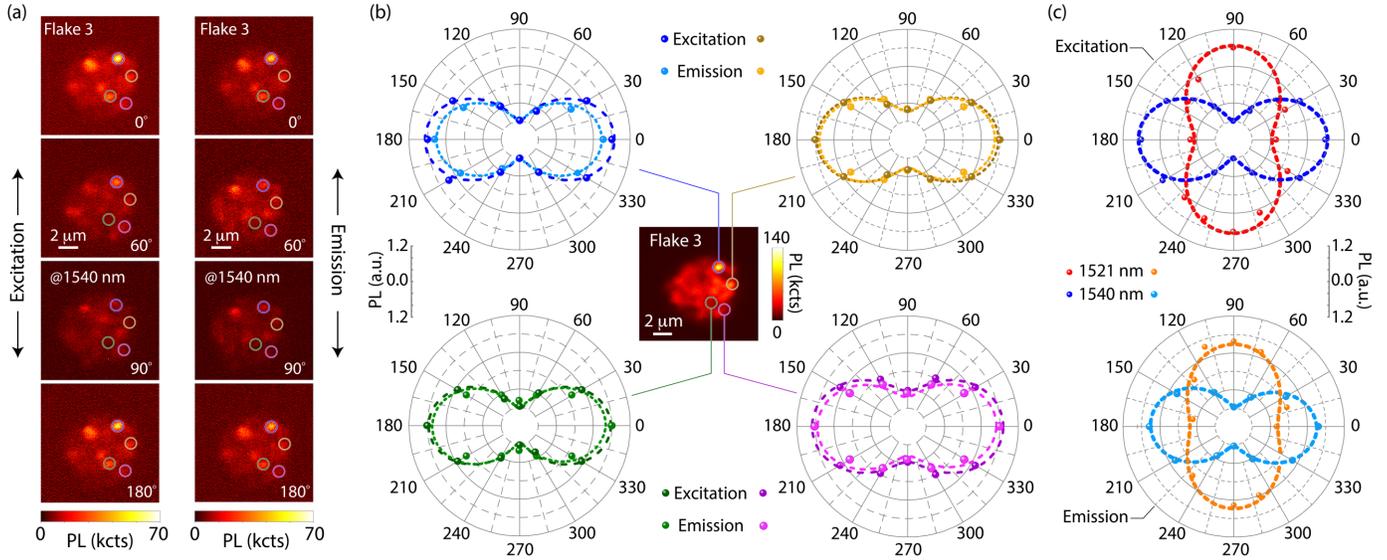

**Figure 4. Polarization dependence of Er$^{3+}$ photoluminescence.** (a) PL images from Flake 3 ($10^{14}$ ions/cm$^2$ at 75 keV) for varying angle of polarization $\theta$ of the excitation beam (left row), or when changing the angle of a linear polarizer prior to PL detection (right row). (b) Excitation and emission polarization plots for four different regions of Flake 3 (colored circles in the PL images); dashed lines in each polar plot are fits to offset sinusoidals. Except the insert image (integrating the PL over the 1500 – 1600 nm range), all measurements limit photon detection to a 10 nm window centered at 1540 nm. (c) Excitation and emission polarization plots (top and bottom, respectively) at 1521 and 1540 nm; in either case we limit photon collection to 10-nm windows. All observations at room temperature.

Our findings suggest a new route to engineering spin-photon interfaces benefiting from the inner-shell transitions of rare-earth emission and the transfer-ready nature of the 2D host. For example, photonic structures may be designed to combine optical resonators and waveguides to simultaneously stimulate telecom emission, efficiently capture outgoing photons, and distribute information to other nodes in a chip; near-field coupling to 2D-crystal-embedded emitters[20] could therefore alleviate constraints otherwise present if the same host material must fulfill these disparate functions.

Future work must examine the Er$^{3+}$ optical and spin coherences as this is an essential element of a spin-photon interface, in turn, critical in applications where the ion serves as a spin qubit or for microwave-to-optical photon transduction[52]. The low natural abundance of all nuclear spin active isotopes portends long spin coherence lifetimes provided the concentration of co-existing paramagnetic impurities can be kept in check. By the same token, one expects long-lived optical coherences if indeed the ion sits at a W site (which is non-polar); since Stark shifts are cancelled only to first order, however, high-quality WS$_2$ flakes featuring low concentrations of dangling bonds and surface charges will be necessary to bring spectral diffusion to a minimum.

A variety of rare-earth ions with transitions in the visible, infrared, and microwave ranges provide opportunities for extending the present results. Besides Er$^{3+}$, other ions of interest include Yb$^{3+}$ (featuring good spin and optical properties[53,54]), Eu$^{3+}$ (seen to attain record spin coherence times[55]), and Ce$^{3+}$ (displaying comparatively high brightness[56]). In the same vein, other 2D hosts are possible including WSe$_2$ and WTe$_2$, immediate extensions where isotopic dilution can bring down the concentration of spin-active nuclei. Of special note is the study of 2D magnetic hosts such as CrBr$_3$[57], where rare-earth ions can serve as local reporters[58].

## AUTHOR INFORMATION

**Corresponding author:**
†E-mail: cmeriles@ccny.cuny.edu

## DATA AVAILABILITY

The data that support the findings of this study are available from the corresponding author upon reasonable request.

## SUPPORTING INFORMATION

Contains details on the experimental setup, measurement protocols and conditions, characterization of WS$_2$ with lower implantation dose, optical spectroscopy and lifetime measurements under cryogenic conditions, and details on ab-initio modeling.

## ACKNOWLEDGEMENTS

G.G.A. and C.A.M. acknowledge support by the U.S. Department of Energy, Office of Science, National Quantum Information Science Research Centers, Co-design Center for Quantum Advantage (C2QA) under contract number DE-SC0012704. G.G.A. also acknowledges support from the Department of Defense Award W911NF-25-1-0134. R.K. and Z.S. acknowledge support from the National Science Foundation via grants NSF-2328993 and NSF-2216838, respectively. C.E.D. acknowledges support from the National Science Foundation under grant NSF-2237674. G.I.L.M acknowledges support from grant NSF-2208863. The authors thank the Surrey Ion Beam Centre staff for carrying out the erbium implantation of the samples. All authors also acknowledge access to the facilities and research infrastructure of the NSF CREST IDEALS, grant number NSF-2112550. The Flatiron Institute is a division of the Simons Foundation.

Supplementary Information for

**Erbium-implanted WS$_2$ flakes with room-temperature photon emission at telecom wavelengths**

Guadalupe García-Arellano[1], Gabriel I. López Morales[1,4], Zav Shotan[1], Raman Kumar[1], Ben Murdin[3], Cyrus E. Dreyer[4,5], and Carlos A. Meriles[1,2,*]

[1]Department of Physics, CUNY-City College of New York, New York, NY 10031, USA.
[2]CUNY-Graduate Center, New York, NY 10016, USA.
[3]Advanced Technology Institute, University of Surrey, Guildford GU2 7XH, United Kingdom.
[4]Department of Physics and Astronomy, Stony Brook University, Stony Brook, New York, 11794-3800, USA.
[5]Center for Computational Quantum Physics, Flatiron Institute, 162 5th Avenue, New York, New York 10010, USA.

I.  **Experimental methods**
    *a) Infrared confocal microscope*
    *b) Sample fabrication.*
    *c) Relation between photoluminescence and flake thickness*
    *d) PL measurements in flakes exposed to varying Er ion doses*
    *e) Optical spectroscopy measurements*
    *f) Optical lifetime measurements*
    *g) Polarization measurements*

II. **Computational methods**
    *a) Quantum Embedding*
    *b) DFT, correlated active space, and computational details*

## I. Experimental Methods

### a) Infrared confocal microscope

Our home-built scanning confocal microscope system features 980-nm excitation lasers and long pass filters to detect emission of $Er^{3+}$ in $WS_2$. The complete system is depicted in Fig.1a of the main text: The key components include two 980-nm excitation lasers, namely, one 1W laser (Optoengine) operating in continuous wave (cw) mode and a 300-mW pulsed laser diode (Thorlabs, HL63163DG) controlled by an IC Haus EVAL HB driver. Each laser is coupled into a single-mode fiber with a 10-μm core size and then collimated with a 50 mm lens. The scanning system uses an XYZ scanning stage (piezo stage Npoint) and an Olympus objective (NA = 0.8, 50x) mounted on a one-axis piezo stage. The beam spot diameter is around 1 micron. We first use incoherent light to localize flakes on the sample using a camera AMScope aligned parallel to the collection path, then we scan a 980-nm beam across a $(100~\mu m)^2$ plane with laser power of 14 mW at a dwell time of 10 ms chosen to attain optimal PL counts. Emission from the samples is collected using the reverse optical path and subsequently separated from the excitation light by a dichroic mirror (Thorlabs DMLP-1138). The PL emission is then focused using a 250-mm magnesium fluoride lens into a single-mode fiber (980 HP, 3.6 μm) and directed into a superconducting nanowire single photon detector (SNSPD 980 IDQ Quantique) optimized to detect telecom wavelengths. We use three filters to prevent excitation laser leakage: one 1200-nm low-pass (LP) filter and two-1500 nm LP filters. Photon counts are recorded using a National Instruments card (NI-PCIe 6321) and a home-made LabVIEW program. Experiments conducted at low temperatures utilize a separate — though similar — confocal microscope [i]. For erbium-based experiments, the system is equipped with two 980 nm excitation lasers: a 100 mW CW laser (MDL III 980 100 mW) and a 300 mW pulsed laser diode (Thorlabs, HL63163DG), driven by an IC Haus EVAL HB. The setup includes a Montana CryoWorkstation (CryoAdvance 100) housing a stack of Attocube positioners (two ANPx101 and one ANPz101) for sample manipulation, and a Zeiss 100x objective with NA 0.90; the same SNSPD is used for telecom photon detection. AFM measurements are performed using a *FastScan* atomic force microscope (AFM) from Bruker.

### b) Sample fabrication

$WS_2$ flakes mechanically exfoliated from a high-purity bulk crystal (2D Semiconductors) were transferred onto a 0.5 mm thick silicon substrate ($4 \times 4~mm^2$). Prior to exfoliation the silicon piece was patterned using UV photolithography to facilitate flake location. As an illustration, Fig. S1 shows an optical image of one region of the sample (0.4 mm x 0.4 mm) where some $WS_2$ flakes are visible.

The samples were broad-area implanted at 75 keV with doses of $10^{12}$, $10^{13}$, and $10^{14}$ ions/$cm^2$ at an angle of 22 deg to avoid channeling. *We respectively denote each flake set as B8, B9, and B10*; all results in the main text correspond to the B10 set but below we present results from the other two sets. Following implantation, all samples were annealed at 400°C in an argon atmosphere for 1 hour (Fig. S1b); the temperature was increased at a rate of 6.3 °C/ minute in a tube furnace (Across International, STF1200).

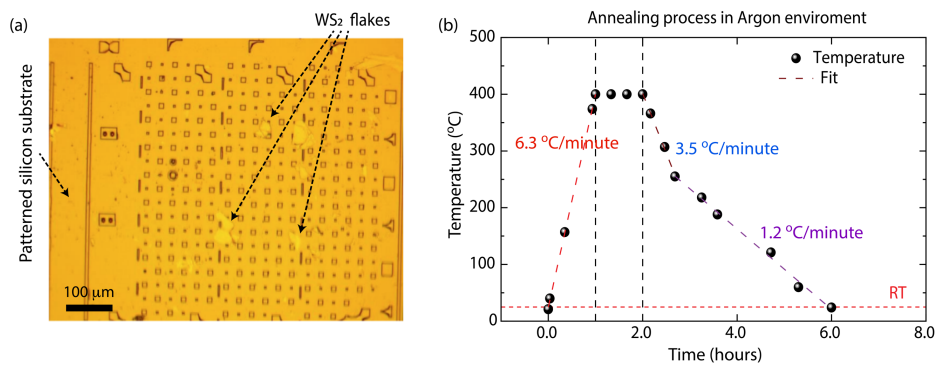

**Fig. S1**: (a) Optical image of a sample region showing various flakes transferred onto a silicon substrate. (b) Annealing protocol

## c) Relation between photoluminescence and flake thickness

Adding to those introduced in the main text, below we present optical, confocal, and AFM images of alternate flakes (Fig. S2). Consistent with the observations in Fig. 1 of the main text, emission can be detected only in regions where the flake thickness is greater than 250 nm.

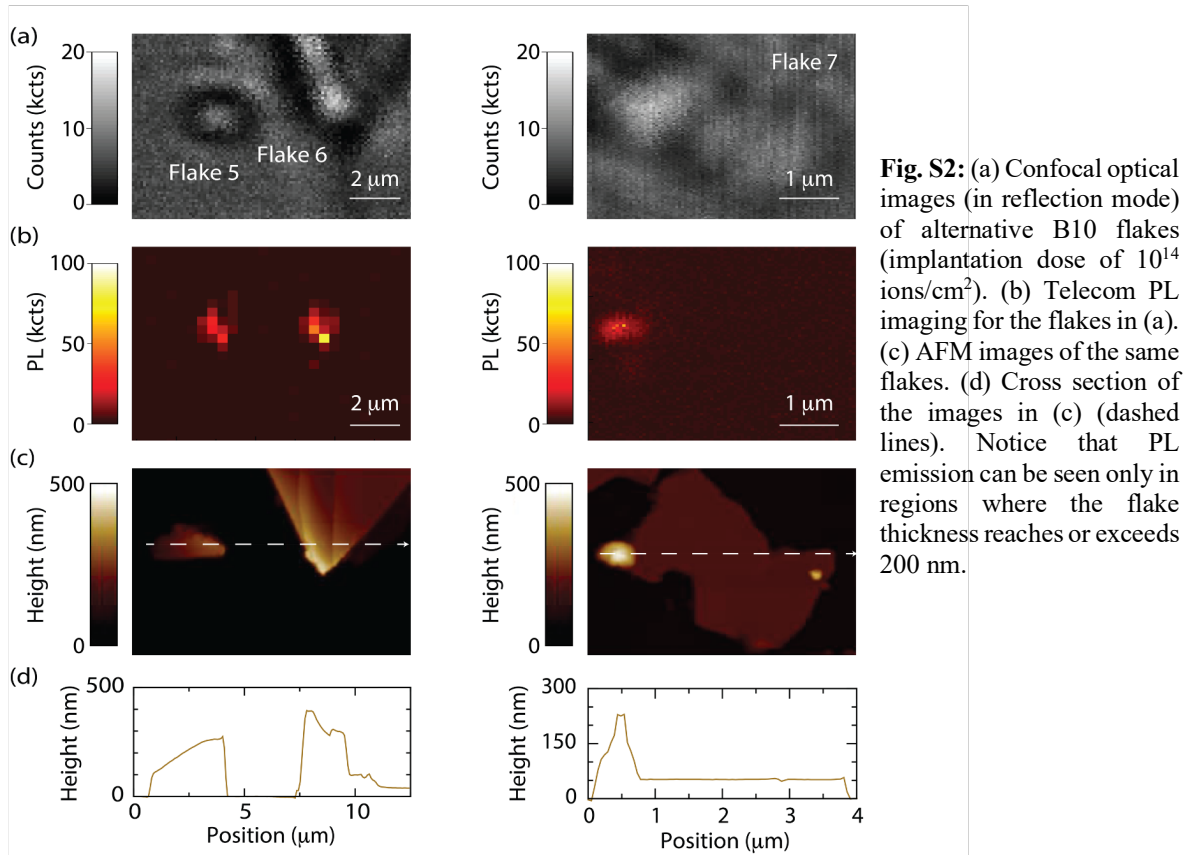

**Fig. S2:** (a) Confocal optical images (in reflection mode) of alternative B10 flakes (implantation dose of $10^{14}$ ions/cm$^2$). (b) Telecom PL imaging for the flakes in (a). (c) AFM images of the same flakes. (d) Cross section of the images in (c) (dashed lines). Notice that PL emission can be seen only in regions where the flake thickness reaches or exceeds 200 nm.

SRIM (Stopping and Range of Ions in Matter) is the traditional software to model ion implantation in bulk materials. However, experimental measurements have shown that it significantly underestimates the penetration depth for ions with atomic number 29 < Z < 83 implanted in thin target materials, mainly due to an incorrect calculation of the electronic stopping powers and the neglect of channeling effects [1-4]. The factor of underestimation can range from 2 (as reported for erbium-implanted silicon films[5]) to 10 (as reported in Xe, Ar, N, and O ions implanted in tungsten[6]).

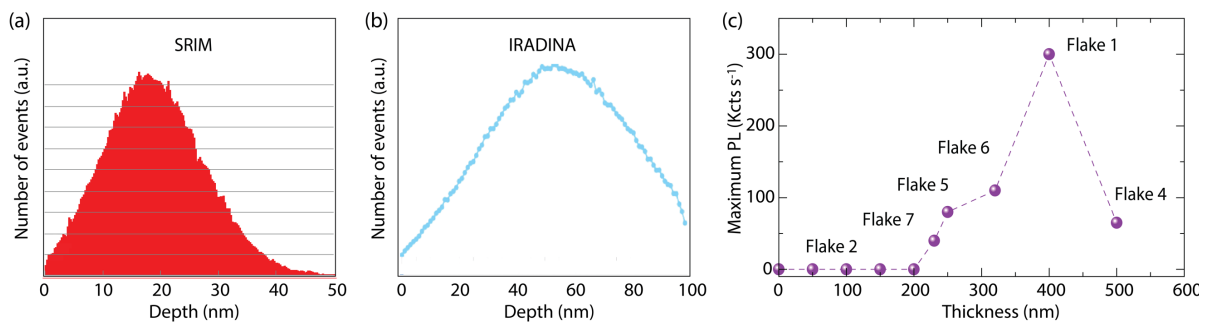

**Fig. S3**: Modeling of the penetration depth of erbium atoms at 75 keV in WS$_2$ calculated by (a) SRIM and (b) IRADINA. (c) Thickness vs maximum luminescence detected in different regions of the erbium implanted WS$_2$ flakes in set B10 (implantation dose of $10^{14}$ ions cm$^{-2}$).

Alternative software based on a binary collision approximation (BCA) and a Monte Carlo (MC) transport algorithm have been developed to estimate with better accuracy the penetration depth; examples are IRADINA[7] or TRI3DYN [8]. In Fig. S3a and S3b we show the penetration depth estimated by different software packages in the case of erbium ions implanted at 75 keV in $WS_2$ at a dose of $10^{14}$ ions/cm$^2$: In the case of SRIM the penetration depth estimated is 25 nm whereas IRADINA predicts 60 nm.

In Figure S3c we plot the maximum PL detected in the B10 flakes and their corresponding thicknesses. We find that flakes thicker than 250 nm exhibit telecom luminescence, indicating that the mean penetration depth of the erbium ions for 75 keV must be comparable to 400 nm. We note that ions of this energy may possibly penetrate even deeper, as the crystallinity of exfoliated flakes deteriorates with increasing thickness, potentially affecting the Er ion fluorescence prematurely (likely the case of Flake 2, in Fig. 1).

*d) PL measurements in flakes exposed to varying Er ion doses*

Besides the B10 set (implantation dose of $10^{14}$ ions/cm$^2$), we also examined other flake sets exposed to a lower ion dose. Figures S4a and S4b present observations in select flakes from sets B9 and B8 (implantation doses of $10^{13}$ and $10^{12}$ ions/cm$^2$, respectively); as in the B10 case, we observe telecom emission in cases where the flake thickness is greater than ~200 nm, confirming the conclusion that the penetration depth of 75 keV ions largely exceeds the SRIM estimates. In some occasions, however, we find non-uniformities in the flake brightness that do not correlate with changes in the flake topography. Examples are the flakes in the first rows of Figs. 4a and 4b where we observe non-fluorescing sections in the PL images that we cannot presently explain; additional work will be needed to clarify this observation.

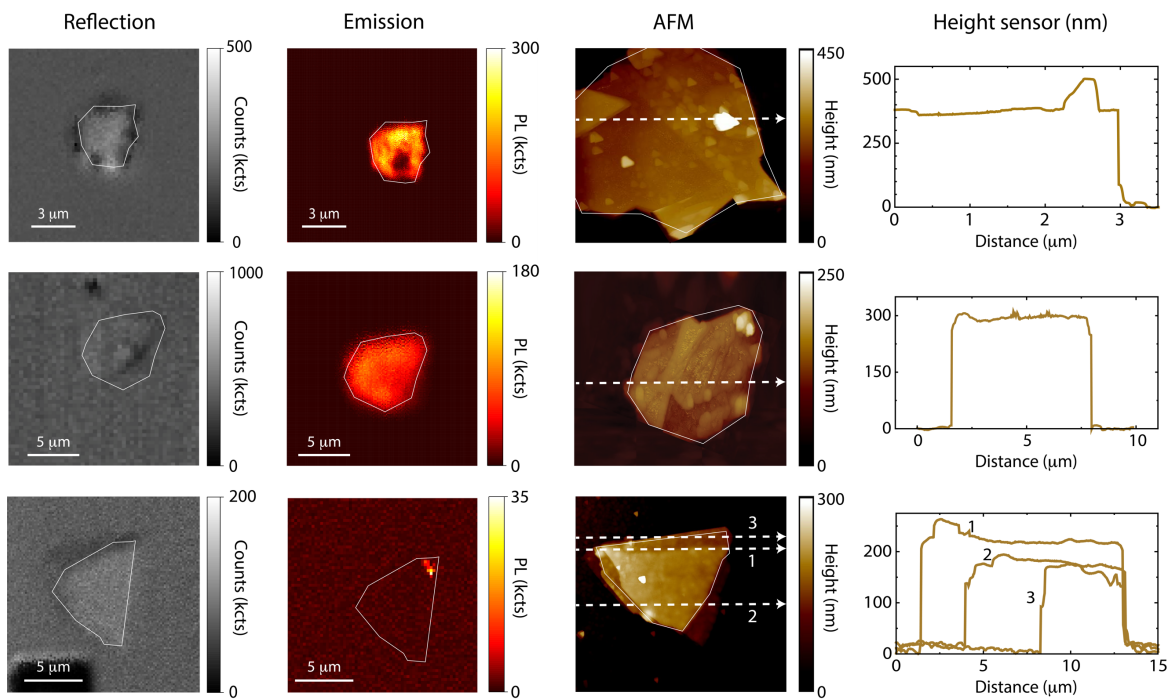

**Fig. S4a**: Optical, confocal, and AFM images (respectively, first, second, and third columns from the left) of select B9 flakes ($10^{13}$ ions/cm$^2$ dose). Plots in the last column show the flake topography across the white dashed line in the corresponding AFM image. As in the B10 set (implantation dose of $10^{14}$ ions/cm$^2$), we see fluorescence in areas where the flake thickness reaches or exceeds 200 nm. All measurements at room temperature. We used a white contour as a guide to the eye. The AFM image in the second row is a composite from partial images of the same flake.

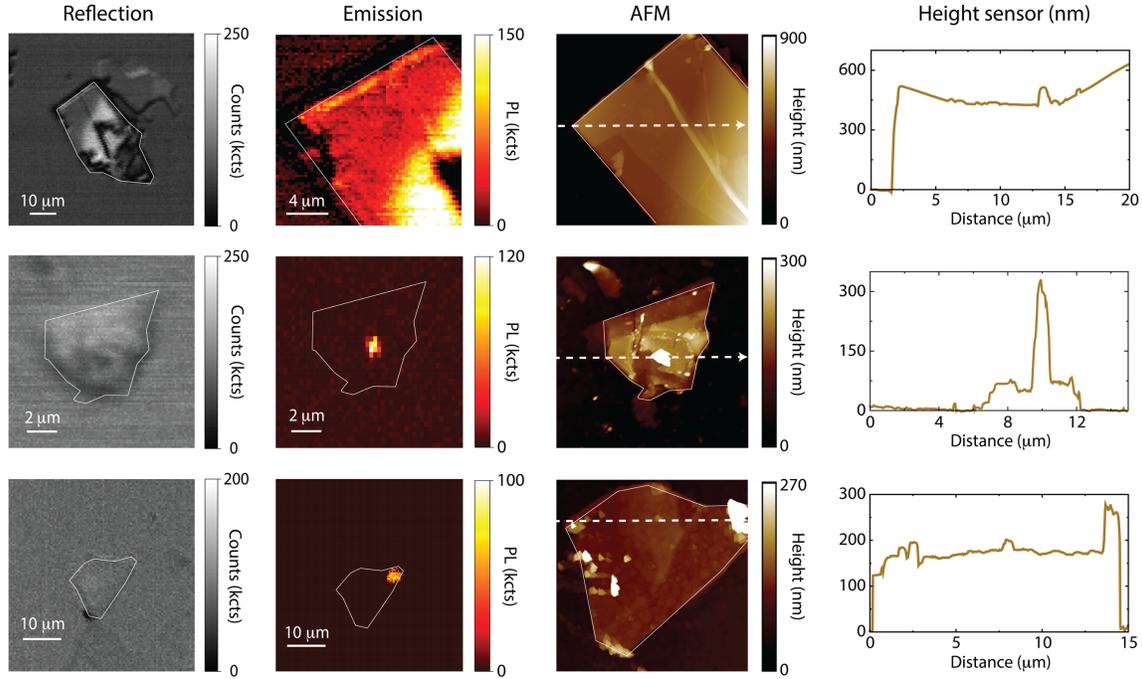

**Fig. S4b**: Same as in Fig. S4a but for some example flakes in the B8 set (implantation dose of $10^{12}$ ions/cm$^2$).

Interestingly, the flakes that do exhibit PL are relatively brighter after accounting for the lower Er content. This is shown in Fig. S5 where we first plot the maximum PL brightness as a function of the WS$_2$ thickness; crucially, we find in all cases that flake sets exposed to a lower Er dose ($10^{12}$ and $10^{13}$ ions/cm$^2$ for sets B8 and B9, respectively) show a PL comparable to that observed in the B10 set (featuring the highest $10^{14}$ ions/cm$^2$ implantation dose). This observation suggests that the conversion efficiency into Er$^{3+}$ is significantly higher for lower Er concentration, although the causes remain unclear: Er has been seen to form large precipitates[9,10] for doses of $4.4 \times 10^{15}$ ions/cm$^2$, a process that could still be at play in the B10 set if annealing can only partially counter atomic aggregation. Consistent with this picture, flakes from the B9 set tend to show a comparatively more uniform brightness (see PL image on the second row of Fig. S4), but additional work will be needed to clarify this point. For completeness, we mention that space charge fields during the implant could potentially lower the ion concentration at the flakes; though possible, this scenario is less likely because it should presumably lead to brightness profiles weaker toward the flake edges, a signature we have not observed.

While determining the ion conversion efficiency necessarily suffers from a large uncertainty, we can leverage the emission rates we measure (see Fig. 2b and below) and the observed PL count at saturation to derive crude estimates of the active Er emitter areal density for each flake set. We derive these values upon considering the collection efficiency of our microscope (detecting approximately 3.5% of the total PL) and the area illuminated by the focused laser (~1 µm$^2$ for our diffraction-limited microscope). Figure S5b shows the results: The conversion efficiency is only 1% for the highest implantation dose (set B10), but grows to reach nearly 30% at lower doses (set B8). We note this value could still improve for optimized annealing recipes, for example, by resorting to slower thermal ramps and/or higher end temperatures.

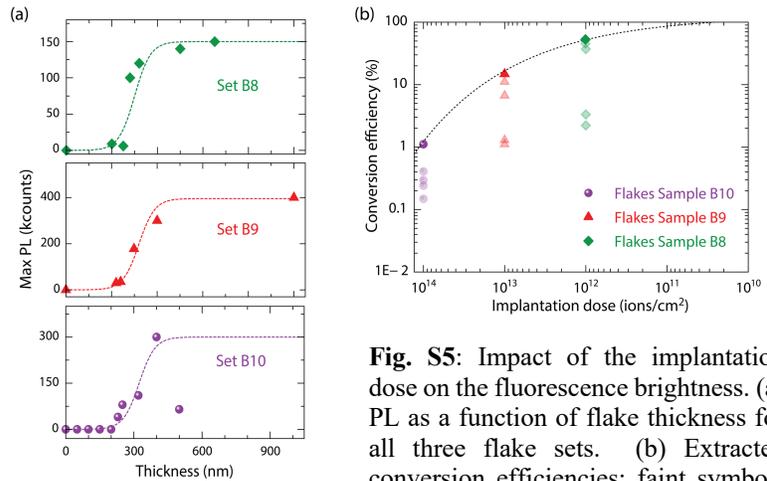

**Fig. S5**: Impact of the implantation dose on the fluorescence brightness. (a) PL as a function of flake thickness for all three flake sets. (b) Extracted conversion efficiencies; faint symbols represent values derived from flakes with suboptimal thickness as presented in (a). The dashed black line is a guide to the eye.

*e) Optical spectroscopy measurements*

Our optical spectroscopy measurements use a Spectra Pro 2300i spectrometer from Princeton Instruments equipped with an InGaAs detector and a 1600 nm grating (600 g/mm blaze) optimized to detect 1.5 µm wavelengths. In our experiments, we first localize the luminescent flake using the confocal microscope, and then park the laser beam in the area of choice. The optical signal is then collimated and sent into the spectrometer. The acquisition time of each spectrum is 60 minutes.

Figure S6 extends the spectroscopy measurements presented in the main text to lower temperatures, down to 3.5 K. Qualitatively, we find the emission spectrum remains largely unchanged, both in terms of the characteristic emission frequencies and relative amplitudes (Fig. S6a), suggesting the coupling with the phonon bath is weak. We do observe, however, some narrowing of the inhomogeneous linewidth, from about 40 to 20 GHz (Fig. S6b). Stemming from the brightest site in Flake 1 of set B10 (Region 2 in Fig. 2 in the main text), these linewidths must be seen as an upper bound. Indeed, measurements in the dimmer sections of the flake (e.g., Region 3) show the room temperature emission linewidth (~12 GHz) is comparable to the spectrometer resolution; notably, no spectra from these regions could be recorded under cryogenic conditions, pointing to linewidths below the detection capability of our spectrometer.

The dominant mechanism driving the observed change in inhomogeneous linewidth is presently unknown, but we hypothesize it may originate from strain fluctuations resulting from Raman and direct phonon scattering by the ion[11], or from changes in the local environment arising from thermal expansion of the WS$_2$ lattice. Using the WS$_2$ Debye temperature of 213 K as a reference[12] and the mild extra broadening observed at room temperature, we surmise the phonon coupling parameters for Er in WS$_2$ must be very small. Similar phonon-induced processes influence the values observed for the transition frequencies at a given temperature[11], implying the above conclusion is consistent with the temperature-insensitive resonances observed in Fig. S6. Additional experiments — including the use of a narrow-band tunable laser to implement high-resolution photoluminescence spectroscopy (PLE) — will be helpful in gaining additional insight, for example, through a careful shape analysis of the absorption peaks[11].

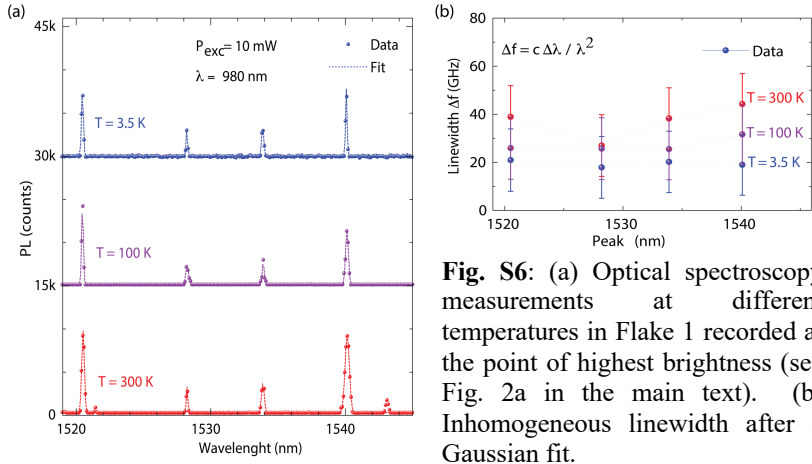

**Fig. S6**: (a) Optical spectroscopy measurements at different temperatures in Flake 1 recorded at the point of highest brightness (see Fig. 2a in the main text). (b) Inhomogeneous linewidth after a Gaussian fit.

*f) Optical lifetime measurements*

We use a pulsed 980 nm diode laser operated by an iC Haus Eval HB driver, in turn controlled via a Pulse Blaster from Spincore. We first obtain a confocal image of the bright flake in the sample, then we park the laser beam at the site of choice. Following a 500-µs-long excitation pulse and a variable wait time, we gate the SNSPD to measure the PL over a 200-µs window; we recover the fluorescence decay profile as we sweep the wait time across a 10-ms interval.

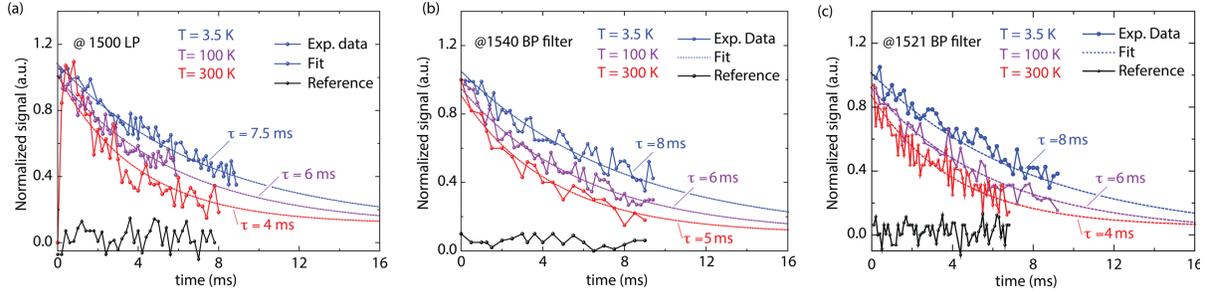

**Fig. S7**: Measuring the excited state lifetime. (a) Excited state relaxation curve at different temperatures as determined from an integrated PL measurement above 1500 nm. (b, c) Same as in (a) but for PL within a 10-nm window centered at 1540 and 1521 nm, respectively. All measurements were carried out on Flake 1 of the B10 set (see Fig. 1 in the main text). LP: Low pass filter. BP: Bandpass filter.

Figure S7 expands the results in Fig. 2b of the main text to include observations at different temperatures and over different spectral windows, namely broadband detection (1500 – 1600 nm, Fig. S7a) or selectively centered at the 1540 and 1521 nm resonances (Figs. S7b and S7c, respectively). We find nominally the same optical lifetimes $\tau$ regardless the spectral window considered, although we do observe a slight increase in $\tau$ with decreasing temperature, from ~4 ms at 300 K to ~8 ms at 3.5 K; the latter is likely due to a suppression of non-radiative relaxation channels under cryogenic conditions. The observed lifetimes are comparable to those reported for Er in other hosts including various oxides[13-17], Si[18-27], garnet materials[28], and fluorophosphate glasses[29,30].

*g) Polarization measurements*

Our experiments integrate a linear polarizer and an achromatic broadband half-wave plate (Thorlabs AHWP05M-980 and AHWP05M-1600) into the excitation and detection paths, respectively. For calibration and to ensure that no artificial effects arise from the optics, we substituted the sample with a fluorescent card (emitting unpolarized light under 980 nm excitation). We then monitored the counts while rotating each of the half-wave plates.

In the excitation path, we also measured the excitation beam power during the rotation of the 980 nm half-wave plate. Emission counts and power data were recorded and are shown in Fig. S8. As illustrated, no systematic variation in either counts or power was observed when rotating the half-wave plates, confirming that the system introduces no artificial polarization effects.

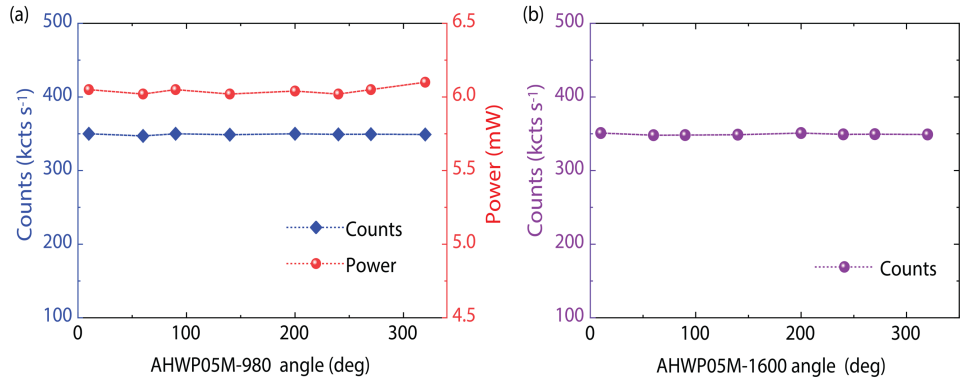

**Fig. S8**: Control measurements on polarization plates. (a) Laser power (red) and fluorescence counts from an unpolarized source (blue) as we change the orientation of a 980-nm half-wave plate controlling the polarization angle of the 980-nm excitation laser beam. (b) Same as in (a) but for the case where we change the orientation of the half-wave plate in the detection path.

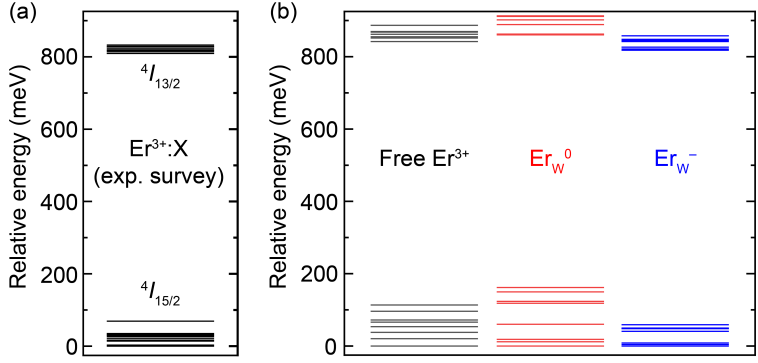

**Fig. S9**: Experimental and calculated many-body spectra. (a) An experimental survey of CF levels for Er in a variety of solid-state materials (Refs. 13 through 30). (b) Calculated many-body spectrum for the isolated $Er^{3+}$ ion in free-space, and the $Er_W$ defect in two different charge states.

## II. Computational methods

### *a) Quantum embedding*

The small spread in optical transitions observed in our experiments suggests a uniform distribution of defect configurations, activated via thermal annealing of the ion-implanted flakes. For this reason, we restrict our simulations to the substitutional Er defect in $WS_2$ ($E_W$), which is also found to be the most stable configuration for $Er^{3+}$ ions in $WS_2$[31]. In order to estimate the optical spectrum of this system, we start by employing density functional theory (DFT) to calculate the relaxed atomic configuration and ground-state electronic structure of a single $Er_W$ defect in monolayer $WS_2$. Due to strong electronic correlations in rare-earth ions, neglected at this level of theory, we use DFT only as a starting point to construct a semi-empirical effective model based on quantum embedding[32-34]. This approach allows us to capture many-body effects between strongly-localized defect states, providing a better description of the emerging optical properties[32-35].

Our effective model takes the form

$$\hat{H}_{eff} = \sum_{\langle ij \rangle} t_{ij} c_i^\dagger c_j + \frac{1}{2} \sum_{\langle ijkl \rangle} U_{ijkl}\, c_i^\dagger c_j^\dagger c_l c_k + \lambda \hat{L}\cdot\hat{S}\,, \qquad (1)$$

where $c_i^\dagger$ ($c$) are creation (annihilation) operators, and $i,j,k,l$ label Er $4f$ states, $t_{ij}$ are hopping matrix elements, $U_{ijkl}$ are screened Coulomb matrix elements, and $\lambda$ is the strength of spin-orbit coupling whose operator is $\hat{L}\cdot\hat{S}$. The spin-orbit operator is constructed using the spin operators $\hat{S}_{x,y,z} = \frac{1}{2}\sum_{i\sigma\sigma'} c_{i\sigma}^\dagger \tau_{\sigma\sigma'}^{x,y,z} c_{i\sigma'}$ ($\tau^{x,y,z}$ are the Pauli matrices) and angular momentum operators $\hat{L}_{z,+,-} = \sum_{ii'\sigma} c_{i\sigma}^\dagger L_{ii'}^{z,+,-} c_{i'\sigma}$ where $L_{ii'}^z = i\delta_{i,i'}, L_{ii'}^+ = \delta_{i,i'+1}\sqrt{l(l+1)-i'(i'+1)}$, and $L_{ii'}^- = \delta_{i,i'+1}\sqrt{l(l+1)-i'(i'-1)}$, where we have used the basis of spherical harmonics ordered by increasing angular momentum quantum number.

We obtain the parameters of the effective model in the following way: The hopping matrix elements are determined from disentangling and Wannierizing the Er $4f$ manifold of states from (non-magnetic) DFT calculations[36] (details below). The direct output from the DFT is symmetrized under the $D_{6h}$ point group. For $U_{ijkl}$, a spherically-symmetric Slater form is assumed with the Hunds coupling parameter $J$ chosen such that for a calculation of an isolated $Er^{3+}$, the spectrum best matches the experiment[37-40]. A similar procedure is used to obtain $\lambda$, where only the quadratic part of the spin-orbit operator is retained.

We obtain the many-body states of the Er defect by diagonalizing Eq. (1), for which we use the Toolbox for Research on Interacting Quantum Systems (TRIQS) software library[41]. In Fig. S9, we compare the crystal-field (CF) levels for $Er^{3+}$ in a variety of solid-state materials[13-30] with the many-body energy diagram obtained through our procedure for the isolated $Er^{3+}$ ion, as well as the neutral and negatively charged $Er_W$ defects. The largest crystal-field (CF) splitting for the $^4I_{15/2}$ manifold of the $Er_W^0$ defect is ~180 meV (note that for the isolated $Er^{3+}$, the CF comes from periodic interactions and is thus arbitrary). For the $Er_W^-$ case, we find a decrease in CF splitting, especially for the $^4I_{15/2}$ manifold. Although DFT is likely to overestimate CF interactions for this system[42], these results suggest that charge compensation around the $Er_W$ defect has a strong impact on the overall CF splitting.

Lastly, we derive the $Er_W$ optical absorption by evaluating the dipole matrix elements across the many-body spectrum using

$$\mu_{ij} = \langle \Psi_i | \hat{r} | \Psi_j \rangle, \qquad (2)$$

where $\Psi_{i,j}$ denote many-body wave functions of states $i$ and $j$, and $\hat{r}$ represents the many-body dipole operator expressed as

$$\hat{r} = \sum_{nm} r_{nm}^{wann}(\mathbf{0}) c_n^\dagger c_m. \qquad (3)$$

The above expression uses the matrix elements of the position operator between Wannier functions $m$ and $n$, given by

$$r_{nm}^{Wann}(\mathbf{R}) = \langle n\mathbf{0} | r | m\mathbf{R} \rangle. \qquad (4)$$

This approach allows for transition dipole matrix elements between explicitly many-body ground and excited states of the $Er_W$ defect, which is important to obtain quantitative transition amplitudes.

For convenience, we fold the optical matrix elements from Eq. (2) onto Lorentzian line-shapes to represent the absorption "strength function" spectrum, i.e.,

$$\tilde{\mu}_{ij} = \sum_{\langle ij \rangle} \frac{\mu_{ij}}{1 + 4\sigma^{-2}(E - E_{ij})^2} \qquad (5)$$

with arbitrary broadening factor $\sigma = 0.10$ meV chosen to resolve closely spaced transitions.

*b) DFT, correlated active space, and computational details*

All DFT calculations are performed using the VASP code[43], with exchange-correlation interactions described via the semi-local generalized-gradient approximation (GGA) parameterized by Pedrew, Burke and Ernzerhof (PBE)[44]. A kinetic energy cutoff of 700 eV is chosen for the plane-wave basis to ensure well-converged electronic structures and properly capture the highly-localized $4f$ electrons studied herein. We make use of projector-augmented wave (PAW) pseudopotentials to treat core electrons[45], with the Er $4f$ electrons treated explicitly within the valence. We model the defect by using a $7 \times 7 \times 1$ supercell containing a single $Er_W$ substitution. The vacuum region included to avoid interlayer (van der Waals) interactions is 20 Å. All defect calculations are employed at the $\Gamma$-point only. We make use of tight convergence criteria for forces (0.001 eV/Å) and electronic iterations ($10^{-8}$ eV) to ensure well-converged structures and wave functions. The ground-state atomic structure of the $Er_W$ defect is obtained via spin-polarized DFT calculations, which we expect to be more realistic than that obtained without spin polarization (non-magnetic).

For all Wannier function calculations, we use the Wannier90 code[36] as interface with VASP. Since the experimental optical transitions are very close to those observed in other Er-doped systems, we assume these to be $4f \leftrightarrow 4f$ transitions and aim at studying the optical transitions only within the $4f$ manifolds of the proposed Er-related defect. Thus, we define the full $4f$ manifold (14 states) as the active space for our quantum embedding calculations, containing 11 electrons (three of which are unpaired) for the case of $Er^{3+}$. The single-particle occupation of $Er^{3+}$ is found to remain unchanged in both the neutral and negatively charged states of $Er_W$. With this in mind, we define a disentanglement energy window of $[-3.15, +0.50]$ eV around the Fermi level[46] to capture important hybridization effects between the Er states and those from the bulk/dangling bonds.